\documentclass[prl,twocolumn,preprintnumbers,superscriptaddress,amsmath,amssymb]{revtex4-1}
\usepackage{graphicx}
\usepackage{subfigure}
\graphicspath{{figures/}}
\usepackage{mathrsfs}
\usepackage{amsfonts}
\usepackage{times}
\usepackage{amsmath}
\usepackage{leftidx}
\usepackage{color}
\usepackage[colorlinks,linkcolor=blue,citecolor=blue]{hyperref}

\newcommand{\Tr}{\operatorname{Tr}}

\usepackage{braket}
\usepackage{mathtools}
\usepackage{physics}

\definecolor{purple}{rgb}{.6,.1,.6}
\definecolor{darkgreen}{rgb}{.1,.6,.1}

\begin{document}

\title{Anomalous Behaviors of Quantum Emitters in Non-Hermitian Baths}
\author{Zongping Gong}
\affiliation{Max-Planck-Institut f\"ur Quantenoptik, Hans-Kopfermann-Stra{\ss}e 1, D-85748 Garching, Germany}
\affiliation{Munich Center for Quantum Science and Technology, Schellingstra{\ss}e 4, 80799 M\"unchen, Germany}
\author{Miguel Bello}
\affiliation{Max-Planck-Institut f\"ur Quantenoptik, Hans-Kopfermann-Stra{\ss}e 1, D-85748 Garching, Germany}
\affiliation{Munich Center for Quantum Science and Technology, Schellingstra{\ss}e 4, 80799 M\"unchen, Germany}
\author{Daniel Malz}
\affiliation{Max-Planck-Institut f\"ur Quantenoptik, Hans-Kopfermann-Stra{\ss}e 1, D-85748 Garching, Germany}
\affiliation{Munich Center for Quantum Science and Technology, Schellingstra{\ss}e 4, 80799 M\"unchen, Germany}
\author{Flore K. Kunst}
\affiliation{Max-Planck-Institut f\"ur Quantenoptik, Hans-Kopfermann-Stra{\ss}e 1, D-85748 Garching, Germany}
\affiliation{Munich Center for Quantum Science and Technology, Schellingstra{\ss}e 4, 80799 M\"unchen, Germany}
\affiliation{Max Planck Institute for the Science of Light, Staudtstra\ss e 2, 91058 Erlangen, Germany}
\date{\today}

\begin{abstract}
Both non-Hermitian systems and the behaviour of emitters coupled to structured baths have been studied intensely in recent years. Here we study the interplay of these paradigmatic settings. 
In a series of examples, we show that a single quantum emitter coupled to a non-Hermitian bath displays a number of unconventional behaviours, many without Hermitian counterpart.
We first consider a unidirectional hopping lattice whose complex dispersion forms a loop. We identify peculiar bound states inside the loop as a manifestation of the non-Hermitian skin effect. In the same setting, emitted photons
may display spatial amplification markedly distinct from free propagation, which can be understood with the help of the generalized Brillouin zone.
We then consider a nearest-neighbor lattice with alternating loss. We find that the long-time emitter decay always follows a power law, which is usually invisible for Hermitian baths.
Our work points toward a rich landscape of anomalous quantum emitter dynamics induced by non-Hermitian baths.
\end{abstract}
\maketitle

\emph{Introduction.}---A central goal in quantum optics is to achieve strong and tunable interactions between atoms and photons at the quantum level. A promising recent strategy is to use nanofabricated lattices in low dimensions, which give rise to structured environments with fundamentally different properties than traditional approaches~\cite{Chang2018}. This has spurred an interest in studying the dynamics of a single or few quantum emitters coupled to structured baths~\cite{Alejandro2017a,Alejandro2017b,Bello2019}. 

Meanwhile, non-Hermitian (NH) physics has become an emergent field with great current interest \cite{Kunst2021,Ashida2021}. This tendency is partially driven by the experimental development in dissipation engineering \cite{Muller2012}, which enables the preparation and control of NH systems in various atomic, molecular and optical platforms. These NH systems enjoy several unique features without Hermitian counterparts, such as the NH skin effect \cite{Yao2018} and exceptional points \cite{Guo2009}. The former refers to the localization of ``bulk modes" and is related to genuine NH topology \cite{Gong2018,Okuma2020,Zhang2020}. The latter refers to the points in the parameter space, 
where the Hamiltonian is not diagonalizable \cite{Heiss2012}. 

\begin{figure}
	\includegraphics[width=8cm, clip]{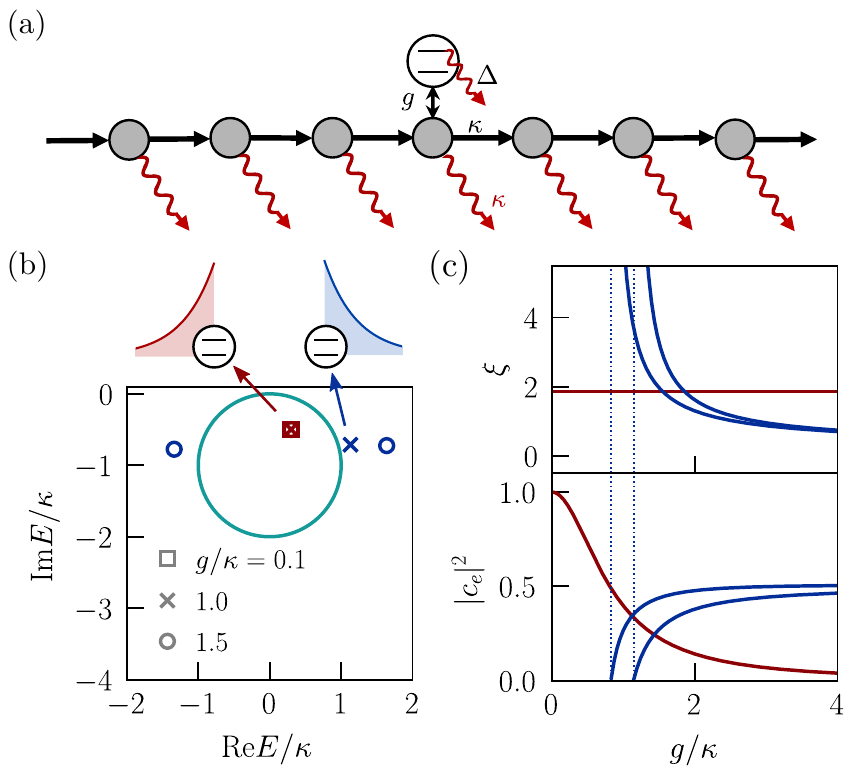}
	\caption{(a) Two-level quantum emitter coupled to a Hatano-Nelson lattice with unidirectional right hopping and background loss. The detuning of the emitter is complex, $\Delta = (0.3 - 0.5 i)\kappa$, including thus purely atomic decay. (b) Single-excitation spectrum under periodic boundary conditions (BCs) for three different coupling strengths. The schematic drawings on top show the profiles of the bound states inside and outside the loop, which are skin-mode- and Hermitian-like (denoted by red and blue markers, respectively). We dub the former ``hidden" bound states, as we will see that they do not apprently affect the emitter decay. Unlike Hermitian-like bound states, the energy of the hidden bound state stays pinned at the emitter detuning, irrespective of the value of $g$. (c) Dependence of the localization length $\xi$ and atomic weight $|c_e|^2$ of the hidden (red) and Hermitian-like (blue) bound states on coupling strength $g$. Dashed vertical lines indicates the onset of the Hermitian-like bound state.}
	\label{fig1}
\end{figure}

\begin{figure}
	\includegraphics[width=8.5cm, clip]{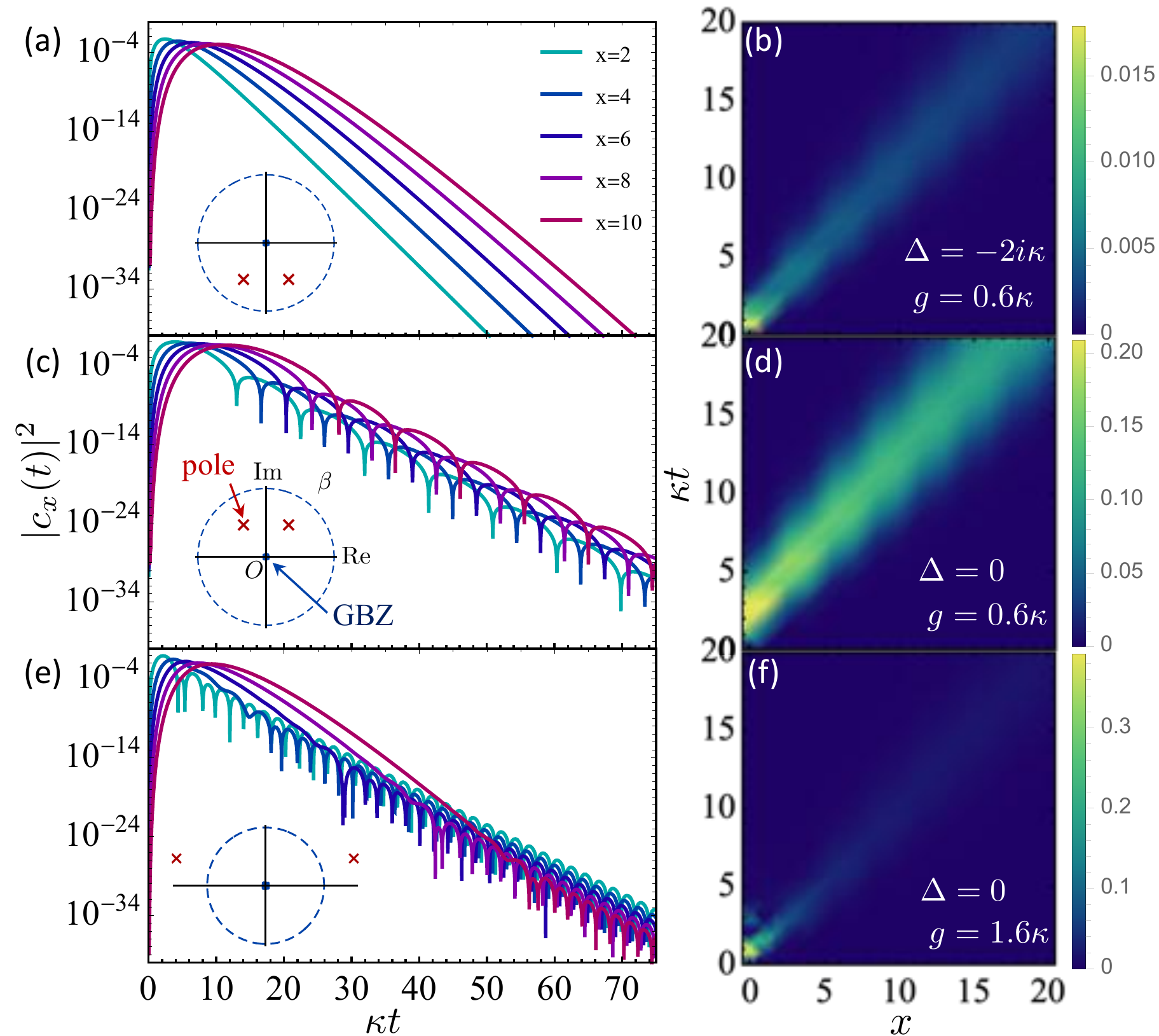}
	\caption{Real-space dynamics of the emitted photon for $\Delta=-2i\kappa$, $g=0.6\kappa$ (a, b), $\Delta=0$, $g=0.6\kappa$ (c, d) and $\Delta=0$, $g=1.6\kappa$ (e, f). Insets in (a), (c) and (e) show the original (dashed) and generalized (solid) Brillouin zone as well as the poles (red crosses) of Eq.~(\ref{cRW}) in terms of $\beta\equiv e^{-ik}$. In (b), (d) and (f), the photon profiles have been normalized by multiplying $\sqrt{\kappa t}$ to retrieve probability conservation asymptotically.}
	\label{fig2}
\end{figure}

\begin{figure}
	\includegraphics[width=8cm, clip]{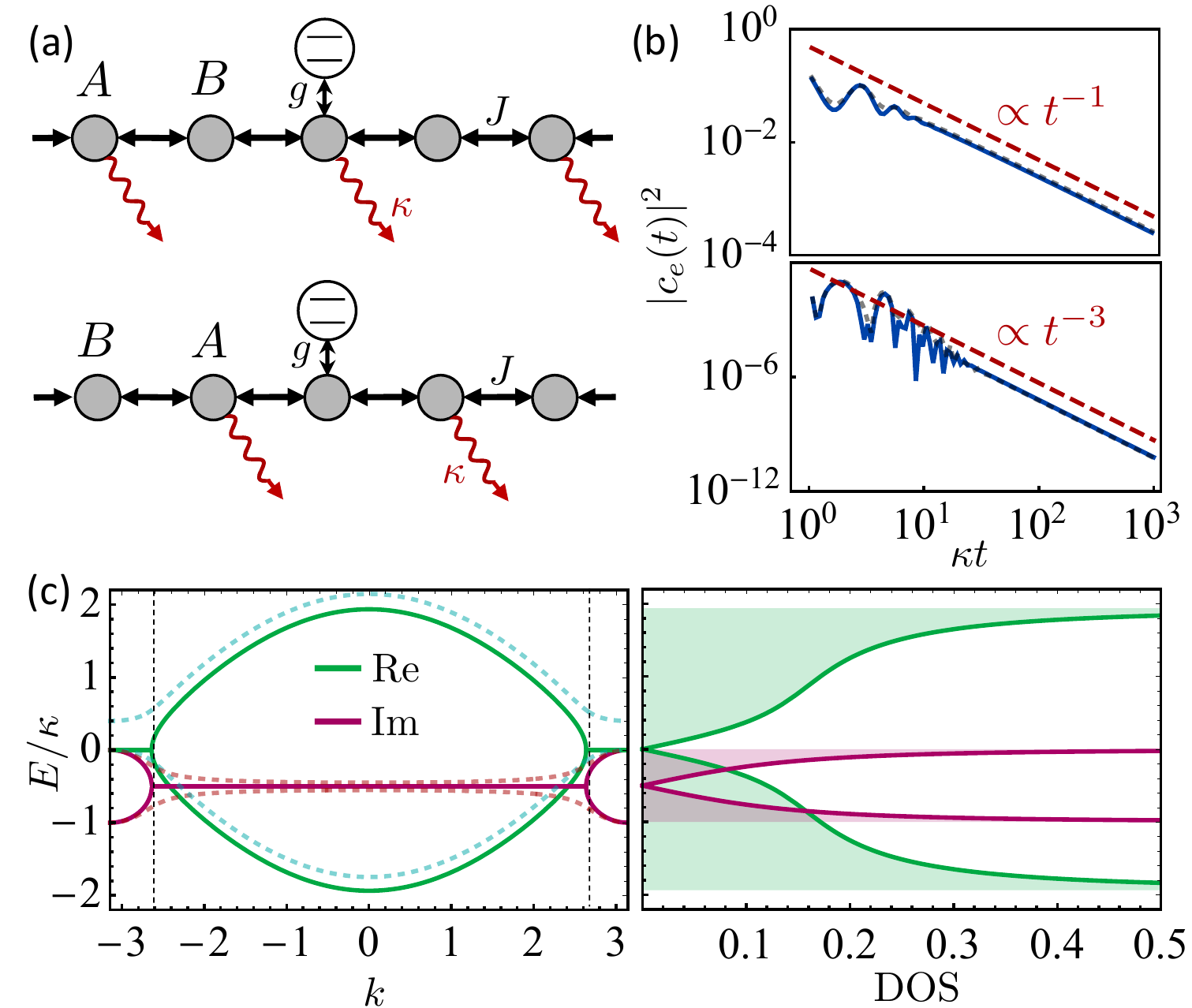}
	\caption{(a) Two-level quantum emitter coupled to a dissipative (upper) or non-dissipative (lower) site of a 1D lattice with symmetric hopping and alternating on-site loss. (b) Asymptotic algebraic decay of the atomic excitation. The power-law behavior is $t^{-1}$ and $t^{-3}$ for the dissipative and non-dissipative coupling, respectively. Here $\Delta=0$, $J=\kappa$ and $g=1.5\kappa$. (c) (left) Band structure and the (right) density of states (DOS) for the alternatively lossy lattice. Dashed vertical lines indicate the exceptional points. Dotted curves in (b) and (c) correspond to a perturbation that explicitly breaks the passive $PT$ symmetry and destroys the EPs.}
	\label{fig3}
\end{figure}

In this Letter, we marry these two fields and point out a few genuinely NH phenomena that emerge already for single quantum emitters coupled to structured lossy baths in one dimension (1D). We focus on three aspects of the quantum-emitter setting: bound states  ~\cite{Tao2016}, photon emission dynamics, and atom decay dynamics \cite{Alejandro2017a,Alejandro2017b,Bello2019}. For the former two aspects, we analyze the Hatano-Nelson \cite{Hatano1996} lattice with nontrivial point-gap topology \cite{Gong2018}. We unveil the existence of ``hidden" bound states with skin-effect origin 
[cf. Fig.~\ref{fig1}(b)] and a diversity of dynamical regimes [cf. Fig.~\ref{fig2}], both without Hermitian counterparts. For the last aspect, we analyze a lattice with passive parity-time ($PT$) symmetry with exceptional points in the band structure. We demonstrate an algebraic asymptotic decay of the atomic excitation [cf. Fig.~\ref{fig3}(b)], which is usually invisible in 1D Hermitian systems.

\emph{Setup.}---We start by considering a quantum emitter, 
modeled as a two-level atom, 
coupled to a 1D nanophotonic lattice with engineered photon loss. The atom may further undergo spontaneous decay from its excited state $|e\rangle$ to the ground state $|g\rangle$. 
Under the Markovian and rotating-wave approximations, 
the equation of motion (in the rotating frame) reads $(\hbar=1)$
\begin{equation}
\dot{\hat\rho}_t=-i[\hat H_{\rm a}+\hat H_{\rm p}+\hat V,\hat \rho_t]
+\mathcal{L}_{\rm a}\hat\rho_t
+\mathcal{L}_{\rm p}\hat\rho_t,
\label{ME}
\end{equation}
where $\hat H_{\rm a}= \Delta_0 \hat\sigma^{ee}$ is the atom Hamiltonian with detuning $\Delta_0$, $ \hat H_{\rm p} = \sum_{x, x'} J_{xx'} \hat a^\dag_x \hat a_{x'}$ is the photon Hamiltonian and $\hat V = g(\hat\sigma^{ge} \hat a^\dagger_{x_0} + {\rm H.c.})$ gives the photon-atom interaction. 
Here, $\hat\sigma^{ww'} \equiv \ket{w}\bra{w'}$ ($w,w' = e,g$), $\hat a_x^\dagger$ ($\hat a_x$) creates (annihilates) a photon at site $x\in\mathbb{Z}$, $g$ is the single-photon Rabi frequency, and $x_0$ is the atom's location. 
The atom and photon dissipators are given by $\mathcal{L}_{\rm a} =\gamma\mathcal{D}[\hat\sigma^{ge}]$ and $\mathcal{L}_{\rm p} = \kappa \sum_x \mathcal{D}[\hat L_x]$,
where $\gamma$ and $\kappa$ control 
the atom decay and photon loss rates, respectively, and $\mathcal{D}[\hat L]\hat\rho \equiv \hat L \hat\rho \hat L^\dagger - \{\hat L^\dagger \hat L,\hat\rho\}/2$ is the Lindblad superoperator. We assume there is only single photon loss, so that $\hat L_x=\sum_{x'} l_{xx'}\hat a_{x'}$ is a linear combination of photon annihilation operators. 

The effective NH Hamiltonian then reads $\hat H_{\rm eff} = \hat H_{\rm a, eff} + \hat H_{\rm p, eff} + \hat V$ with $\hat H_{\rm a,eff}= (\Delta_0-\frac{i}{2}\gamma)\hat\sigma^{ee}$ and $\hat H_{\rm p, eff} = \hat H_{\rm p} - \frac{i}{2} \kappa \sum_x \hat L^\dagger_x \hat L_x$, the latter of which is quadratic under the above assumption. 
Suppose the initial state is in the single-excitation sector, such as 
$\hat\rho_0 = \ket{\psi_0}\bra{\psi_0}$ with $\ket{\psi_0} = \hat\sigma^{eg} \ket{g} \otimes \ket{\rm vac}$ and $\ket{\rm vac}$ being the photon vacuum, then the solution to 
Eq.~(\ref{ME}) reads 
$\hat\rho_t= e^{-i\hat H_{\rm eff} t}\hat\rho_0e^{i\hat H^\dag_{\rm eff} t}+ p_t|g\rangle\langle g|\otimes|{\rm vac}\rangle\langle {\rm vac}|$ 
with $p_t=1-\Tr[e^{-i\hat H_{\rm eff} t}\hat\rho_0e^{i\hat H^\dag_{\rm eff} t}]$ \cite{Alejandro2017b,Gong2017}. Therefore, as long as we restrict ourselves to the single-excitation sector, we may focus on studying the effective NH Hamiltonian. For translation-invariant lattices with $J_{xx'}=J_{x-x'}$, $l_{xx'}=l_{x-x'}$ and $x_0=0$, the effective NH Hamiltonian takes the general form
\begin{equation}
    \hat H_{\rm eff} = \Delta \hat\sigma^{ee} + \sum_k \omega_k \hat a^\dagger_k \hat a_k + \frac{g}{\sqrt{L}} \sum_k (\hat \sigma^{ge} \hat a_k^\dagger + {\rm H.c.}), 
\label{eq:eff_ham_mom_space}
\end{equation}
where $\Delta=\Delta_0 -\frac{i}{2}\gamma$, $L$ is the lattice length, $\hat a_k = L^{-1/2} \sum_x e^{- i k x} \hat a_x$ and $\omega_k= J_k -\frac{i}{2}\kappa |l_k|^2$, with $J_k=\sum_x J_xe^{-ikx}$ and $l_k=\sum_x l_xe^{-ikx}$, is the complex band dispersion of $\hat H_{\rm p,eff}$. 

\emph{Bound state.}---The Hermitian version of Eq.~(\ref{eq:eff_ham_mom_space}) has been widely studied in the literature. Of particular interest are photon--atom bound states, which are eigenstates of Eq.~(\ref{eq:eff_ham_mom_space}) with photon profiles localized around the atom
\cite{John1990}. To find the bound state in the NH case, we can follow exactly the same procedure as the Hermitian case: we first write down a general state 
$\ket{\psi_{\rm b}} = \left[c_e \hat\sigma^{eg} + L^{-1/2} \sum_k c_k \hat a_k^\dagger \right] \ket{g} \otimes \ket{\rm vac}$ in the single-excitation sector. 
Imposing the eigenstate condition $\hat H_{\rm eff} \ket{\psi_{\rm b}} = E_{\rm b} \ket{\psi_{\rm b}}$, we obtain
        $\Delta c_e + \frac{g}{L} \sum_k c_k = E_{\rm b} c_e$ and $\omega_k c_k + g c_e = E_{\rm b} c_k$,  $\forall k$.
Solving these equations and using that a bound state should have $c_e \neq 0$, we find 
\begin{equation}
    E_{\rm b} - \Delta -\Sigma(E_{\rm b})=0,\;\;\;\;\Sigma(E)=\frac{g^2}{L} \sum_k \frac{1}{E - \omega_k},
    \label{Eb}
\end{equation}
where $\Sigma(E)$ is the self-energy of the quantum emitter. Then, we obtain
$|c_e|^2=(1+g^2L^{-1}\sum_k|E_{\rm b}-\omega_k|^{-2})^{-1}$
and $c_k=g(E_{\rm b}-\omega_k)^{-1} c_e$, whose (inverse) Fourier transform gives the real-space photon profile. Note that the formula $|c_e|^2=(1-\Sigma'(E_{\rm b}))^{-1}$ for Hermitian systems breaks down in general due to the fact that both $E_{\rm b}$ and $\omega_k$ may be complex. 

A unique feature of 1D NH systems is that the spectrum under periodic BCs can form a loop, which is characterized by the spectral winding number \cite{Gong2018}. This is a signature of NH topology and has been identified as the origin of the NH skin effect \cite{Okuma2020,Zhang2020}. A prototypical model that exhibits such nontrivial spectral winding is the Hatano-Nelson model with asymmetric hopping \cite{Hatano1996}. Choosing $\hat L_x = \hat a_x - i \hat a_{x+1}$ \cite{Gong2018}
and $J_x=\kappa(\delta_{x,1}+\delta_{x,-1})/2$, 
we can even reach the maximally NH limit with unidirectional (right) hopping [cf. Fig.~\ref{fig1}(a)], in which case $\omega_k=\kappa (e^{-ik}-i)$ under periodic BCs and thus the spectrum forms a loop $|E+i\kappa|=\kappa$ [cf.  Fig.~\ref{fig1}(b)]. Substituting the expression of $\omega_k$ into Eq.~(\ref{Eb}) and taking the thermodynamic limit $L\to\infty$, one finds
\begin{equation}
    \Sigma(E)=\left\{\begin{array}{ll} 0, & |E+i\kappa|<\kappa,\\ \frac{g^2}{E+i\kappa}, & |E+i\kappa|>\kappa. \end{array}\right.
\end{equation}
It turns out that the self-energy vanishes 
inside the loop. This result actually has a topological origin \cite{Gong2022} and implies the existence of a bound state ``hidden" in the loop, with its energy pinned at $\Delta$. Its photon profile, obtained by (inverse) Fourier transforming $c_k$, vanishes to the right of the emitter and decays exponentially to the left, with localization length $\xi=(\ln|\kappa/(\Delta + i\kappa)|)^{-1}$. Notably, it does not depend on the coupling strength $g$. In contrast, bound states with energies outside the loop only arise for sufficiently large $g$, and feature a photon profile that decays exponentially in the right-half space with localization length strongly depending on $g$. See Figs.~\ref{fig1}(b) and (c) for an illustration, 
where one also finds an opposite $g$-dependence of the atom weight $|c_e|^2$ for these two different types of bound states. 

The behavior of the hidden bound state 
is reminiscent of the vacancy-like bound state in Hermitian topological lattices \cite{Luca2021}, such as the Su-Schrieffer-Heeger model \cite{Bello2019}. Indeed, there is an analogous interpretation: removing the site to which the atom is coupled, the resulting systems has a skin mode with energy $\Delta$, as long as $\Delta$ is within the loop \cite{Reichel1992,Gong2018,Okuma2020,Zhang2020}. A bound state can then be obtained as a proper superposition between the skin mode and the atomic excitation. Note that in stark contrast to the Hermitian case, there is no need to fine-tune $\Delta$ since 
skin modes form a continuum. The bound states outside the loop behave more like conventional Hermitian bound states. Nevertheless, unlike the Hermitian case, where bound states appear for arbitrarily small $g$ in 1D \cite{Tao2016}, here we need a sufficiently large $g$. 
This is closely related to the fact that the Anderson transition in the Hatano-Nelson model occurs at finite disorder strength~\cite{Hatano1996}, whereas 1D Hermitian models immediately localize \cite{Abrahams1979}. Here an emitter as a specific type of disorder requires finite $g$ to localize extensive modes in the Hatano-Nelson model to form a bound state.

While 
beyond the scope of this short letter, we mention that some properties of the hidden bound states are inherited by the more general Hatano-Nelson model (
with 
nonzero left and right hopping) even in the multiple emitter case. For example, the eigenenergies are always pinned at the (generally different) complex detunings, as if the emitters did not influence each other. Further details can be found in the companion paper \cite{Gong2022}, where we also discuss the effect of the BCs.

\emph{Dynamics.}---We move on to study the photon emission dynamics, for which the BCs are irrelevant \cite{Mao2021}. To be specific, we focus on the non-unitary evolution $|\psi_t\rangle=e^{-i\hat H_{\rm eff}t}|\psi_0\rangle$ starting from an excited atom $|\psi_0\rangle=|e\rangle\otimes|{\rm vac}\rangle$, an initial state that can be easily realized in experiments \cite{Cai2019}. Expanding the time-evolved state as $|\psi_t\rangle=[c_e(t) \hat\sigma^{eg} + L^{-1/2}\sum_k c_k(t) \hat a_k^\dagger ] \ket{g} \otimes \ket{\rm vac}$ ($c_{\alpha}(0)=\delta_{\alpha e}$, $\alpha=e,k$) without loss of generality, the coefficients can be calculated as 
\begin{equation}
c_\alpha(t)=\frac{i}{2\pi}\int^\infty_{-\infty} dE\, G_\alpha(E+i0^+)e^{-iEt} 
\label{cat}
\end{equation}
using the resolvent method \cite{Alejandro2017a,Tannoudji1998}. Here the emitter ($\alpha=e$) and photon ($\alpha=k$) Green's functions are given by
\begin{equation}
G_e(E)=\frac{1}{E-\Delta-\Sigma(E)},\;\;\;\;
G_k(E)=\frac{gG_e(E)}{E-\omega_k},
\label{GF}
\end{equation}
where 
$\Sigma(E)$ follows that in Eq.~(\ref{Eb}). Note that the integral in Eq.~(\ref{cat}) is performed above the real axis. For the Hatano-Nelson model, this implies that the bound states outside the loop are picked as poles, while that inside 
has no apparent contribution \cite{Longhi2016}. 
This is the main reason why we call the latter 
``hidden", and is consistent with its skin-effect origin since skin modes are \emph{bulk} modes associated with the unstable poles (i.e., poles of $G_e$ with $\Sigma$ analytically continued from the exterior of the loop). This point also serves as a crucial difference from vacancy-like bound states in Hermitian systems \cite{Luca2021}, which do contribute to the dynamics as regular poles. 


Thanks to the simplicity of the unidirectional Hatano-Nelson model, we can obtain the analytic expressions for both emitter and photon dynamics by substituting $\Sigma(E)=g^2/(E+i\kappa)$ and $\omega_k=\kappa(e^{-ik}-i)$ into Eqs.~(\ref{cat}) and (\ref{GF}). The emitter decay is normally 
exponential, possibly with 
oscillations or polynomial corrections. For example, taking $\Delta=0$, we obtain a decay rate $(\kappa-\sqrt{\kappa^2-4g^2})/2$ ($\kappa/2$) for $g<\kappa/2$ ($g>\kappa/2$), which asymptotically reads $g^2/\kappa$ for $g\ll\kappa$, as a manifestation of the continuous Zeno effect \cite{Knight2000}. 

Unlike the exponential emitter decay, the overall photon population $\lim_{L\to\infty}L^{-1}\sum_k|c_k(t)|^2=\int^\pi_{-\pi}\frac{dk}{2\pi}|c_k(t)|^2$ decays asymptotically as $t^{-1/2}$ due to the vanishing damping gap at $k=-\pi/2$, as is understandable by evaluating $\int^\pi_{-\pi} dk e^{2{\rm Im}\omega_k t}$ \cite{Song2019}. The real-space dynamics of the emitted photon 
can be solved exactly by Fourier transforming $c_k(t)$ as
\begin{equation}
c_x(t)=\frac{ge^{-\kappa t}[\zeta_+^{-x}R_x(-i\kappa\zeta_+t)- \zeta_-^{-x}R_x(-i\kappa\zeta_-t)]}{\kappa(\zeta_+-\zeta_-)},
\label{cjt}
\end{equation}
where $x\ge0$ (otherwise $c_x(t)=0$), 
$\zeta_\pm =(E_\pm+i\kappa) /\kappa$ with $E_\pm=[(\Delta-i\kappa)\pm\sqrt{(\Delta+i\kappa)^2+4g^2}]/2$ being the poles of $G_e(E)$ (with $\Sigma(E)$ always taken as $g^2/(E+i\kappa)$) and $R_x(z)= e^z-(\sum^x_{n=0} z^n/n!)$.  
One can identify several different dynamical regimes from Eq.~(\ref{cjt}): If ${\rm Im}\zeta_\pm\le0$, the long-time asymptotic behavior reads $c_x(t)\simeq -\kappa e^{-\kappa t}(-i\kappa t)^{x-1}/[g(x-1)!]$ for $x\ge1$, which resembles very much the free photon propagation. Otherwise, the asymptotic expression is given by $c_x(t)\simeq ge^{-\kappa t}(\zeta_+^{-x}e^{-i\kappa\zeta_+t} - \zeta_-^{-x}e^{-i\kappa\zeta_-t})/[\kappa(\zeta_+ - \zeta_-)]$, which implies a spatial amplification (decay) for $|\zeta_\pm|<1$ ($|\zeta_\pm|>1$). Typical plots for all these regimes are shown in Fig.~\ref{fig2}. 

The last dynamical regime is easiest to understand and is common in Hermitian systems: We eventually observe the profile of the bound state, which decays in space. The first two regimes require the absence (or sufficiently rapid decay) of the bound states (outside the loop), and are thus unique to NH systems. To determine whether the spatial amplification is free-like or emitter-dependent, we can borrow the wisdom of the generalized Brillouin zone \cite{Yao2018}. Excluding the bound-state and possible branch cut contributions, the running wave contribution in real space reads \cite{Alejandro2017b}
\begin{equation}
c^{\rm RW}_x(t) = g\int^\pi_{-\pi} \frac{dk}{2\pi} \frac{e^{-i\omega_k t+ ikx}}{\omega_k-\Delta-\Sigma(\omega_k+i\eta_k)},
\label{cRW}
\end{equation}
where $\eta_k$ is an infinitesimal quantity such that $\omega_k+i\eta_k$ lies right outside the complex spectrum, where $\Sigma(E)$ is analytic. To perform the stationary-phase approximation \cite{Bender1999}, we have to deform the integral contour from the conventional Brillouin zone to the generalized one, during which the contour may cross one or several poles $\tilde k$, leading to additional terms with spatial amplification determined by ${\rm Im}\tilde k$ and temporal decay determined by ${\rm Im}\omega_{\tilde k}$. In our specific example, the poles are nothing but $\zeta_\pm$. If ${\rm Im}\zeta_\pm >0$ (${\rm Im}\zeta_\pm <0$), which means ${\rm Im}\tilde k$ is larger (smaller) than the imaginary energy of the running-wave evaluated from the generalized Brillouin zone (which is simply $\beta\equiv e^{-ik}=0$; cf. Fig.~\ref{fig2}(c)), the amplification governed by the poles (free-like propagation) will dominate the long-time behavior. We mention that the three dynamical regimes appear also in the general Hatano-Nelson model and a similar analysis applies \cite{Gong2022}.



\emph{Critical emitter decay.}---While the emitter decay into the Hatano-Nelson bath  is trivially exponential, we show that novel critical (algebraic) decay emerges for another simple NH bath with alternating on-site loss. Here the criticality may be considered as a temporal analogy of the power-law spatial decay of correlations in conventional critical phases \cite{Sachdev2011}. Just like the spatial case, the criticality is related to the gaplessness of the underlying system, but along the imaginary axis.

As illustrated in Fig.~\ref{fig3}(a), with the sublattices consisting of even (odd) sites denoted as $A$ ($B$), this model can be readily constructed by taking $\hat L_x=\sqrt{(-1)^x+1}\hat a_x$ and the same nearest-neighbor Hamiltonian $\hat H_{\rm p}=J\sum_x(\hat a_{x+1}^\dag\hat a_x +{\rm H.c.})$. Suppose that the total length of the lattice is $2L$, we can define $\hat{\boldsymbol{a}}_k=[\hat a_{kA},\hat a_{kB}]^{\rm T}$ with $\hat a_{kA}= L^{-1/2}\sum_x e^{-ikx}\hat a_{2x}$ and $\hat a_{kB}= L^{-1/2}\sum_x e^{-ikx}\hat a_{2x+1}$ and write down a similar effective NH Hamiltonian as Eq.~(\ref{eq:eff_ham_mom_space}), except for that $\omega_k \hat a^\dag_k \hat a_k$ should be replaced by $\hat{\boldsymbol{a}}_k^\dag h_k \hat{\boldsymbol{a}}_k$ with
\begin{equation}
h_k = J(1+\cos k)\sigma^x+ J\sin k\sigma^y-\frac{i}{2}\kappa(\sigma^z+\sigma^0),
\label{PThk}
\end{equation}
and that $\hat a_k$ in the coupling (third) term should be specified as $\hat a_{kA}$ or $\hat a_{kB}$, depending on whether the atom is located at a dissipative site or not. We numerically compute the atom decay dynamics governed by this NH Hamiltonian starting from $|\psi_0\rangle=|e\rangle\otimes|{\rm vac}\rangle$. As shown in Fig.~\ref{fig3}(b), we find that the decay asymptotically follows a power law $t^{-1}$ ($t^{-3}$) for the coupling to a dissipative (non-dissipative) site. We mention that $t^{-1}$ requires fine-tuning $\Delta$ to be zero, as is adopted in Fig.~\ref{fig3}(b); otherwise, we observe a $t^{-3}$ decay after a $t^{-1}$ transient \cite{Gong2022}. 
Having in mind the exponential decay in the Hermitian limit 
\footnote{Rigorously speaking, in the Hermitian limit, there is still a small, yet nonzero contribution of algebraic decay, which is nevertheless invisible in the long-time limit due to the bound state. It is thus fair to say that the exponential decay dominates during a physically relevant time, especially for $\Delta=0$ in which case the single-pole approximation, i.e., the Fermi's golden rule, works best \cite{Alejandro2017b}.}, we may call this phenomenon critical Zeno effect, in the sense that adding dissipation qualitatively suppresses the decay to an algebraic one.

To understand the critical decay, we first note that the NH two-band Bloch Hamiltonian in Eq.~(\ref{PThk}) exhibits a passive $PT$-symmetry \cite{Guo2009}, which becomes exact after neglecting the background loss $-i\kappa\sigma^0/2$. This passive $PT$ choice arises naturally from our loss-only setup, yet it can still accommodate genuine NH objects such as EPs. The band dispersions read $-i\kappa/2\pm\sqrt{2J^2(1+\cos k)-\kappa^2/4}$. For not-too-strong loss rate $\kappa<4J$, there are two EPs 
at $k_{\rm EP}=\pm \arccos(\kappa^2/(8J^2)-1)$ in the Brillouin zone. Since these two EPs are smoothly connected over the passive $PT$-broken region, the eigenenergy with maximal imaginary part, which is zero here, should necessarily have divergent density of state [cf. Fig.~\ref{fig3}(c)]. This divergence manifests itself as a branch point in the Green's function (\ref{GF}), giving rise to an algebraic decay. See Ref.~\cite{Gong2022} for a quantitative analysis. Recalling that the branch point has the largest imaginary part, we know that the algebraic decay dominates the long-time dynamics. 

Finally, we would like to clarify that the (passive) $PT$-symmetry breaking and the associated EPs are more like a recipe rather than necessity for observing the critical decay, which indeed survives asymmetric perturbations [cf. Fig.~\ref{fig3}(b) and (c)]. What is important is to have a branch point on the real axis while all the poles well below the real axis. For example, this can be realized by Wick rotating (i.e., multiplying by $i$) the Hermitian bath with nearest-neighbor hopping \cite{Gong2022}. Note that algebraic decay is usually invisible for 1D periodic Hermitian baths \cite{Alejandro2017b}, since there are always bound states with real energies, which dominate the long-time dynamics \footnote{If we do not impose periodicity, then algebraic decay already appears for an emitter coupled to the end of a semi-infinite chain \cite{Longhi2006}. One can check that there is no bound state if the coupling strength $g$ is small enough.}.

\emph{Conclusion and outlook.---}In summary, we have studied the bound states, photon emission and excited-state decay of single quantum emitters in 1D NH nanophotonic lattices. For the Hatano-Nelson lattice, we have found hitherto unknown hidden bound states originating from the skin effect, and unconventional dynamical regimes of photon propagation without Hermitian counterpart. For the alternatingly lossy lattice, we found critical emitter decay in the long-time limit, which is again hindered in Hermitian systems by stable bound states.

Our work opens up a plethora of possibilities for future studies. One big direction is to examine to what extent these NH phenomena survive on the many-body level \cite{Tao2016,Asenjo2017}, where we have to deal with the full Lindblad dynamics \cite{Daley2014b,Weimer2021}. This is particularly relevant in the ultrastrong-coupling regime \cite{Nori2019,Solano2019,Eduardo2019}, where the counterrotating terms become relevant and the conservation of excitation number breaks down. Even on the single-particle level, we can consider dissipative lattices with more complicated NH topology \cite{Lee2019a,Kawabata2019a}, such as class AII$^\dag$ in 1D exhibiting the $\mathbb{Z}_2$ skin effect \cite{Okuma2020} and that in 3D mimicking the edge physics of the 4D quantum Hall effect \cite{Kunst2020}. It may also be interesting to explore the impact of higher-order \cite{Hodaei2017} and higher-dimensional exceptional objects, such as exceptional rings \cite{Xu2017} and surfaces \cite{Zhou2019}, whose systematic constructions have become clear recently \cite{Mandal2021,Delplace2021,Kunst2022}. Last but not least, introducing non-Hermiticity to nanophotonic baths may significantly extend the freedom of engineering effective (bath-mediated) interactions between atoms \cite{Roccati2021}.

\emph{Acknowledgments.---} We acknowledge Ignacio Cirac and Yuto Ashida for helpful discussions. Z.G.\ is supported by the Max-Planck-Harvard Research Center for Quantum Optics (MPHQ). F.K.K.\ was supported by the Max-Planck-Harvard Research Center for Quantum Optics (MPHQ) before moving to MPL. M.B.\ and D.M.\ acknowledge funding from the ERC Advanced Grant QUENOCOBA under the EU Horizon 2020 program (Grant Agreement No. 742102).

\bibliography{GZP_references}

\end{document}